\documentclass[aps, prd, twocolumn, superscriptaddress, nofootinbib]{revtex4}

\usepackage{amsmath, amssymb, bm}
\usepackage{graphicx}
\usepackage{epstopdf}
\usepackage{amsfonts}
\usepackage{amssymb}
\usepackage{amsbsy}
\usepackage{amsmath}
\usepackage{latexsym}
\usepackage{sansmath}
\usepackage{subfigure}
\usepackage{lipsum}
\usepackage{float}
\usepackage{cancel}

\usepackage{bm}
\usepackage{xcolor}

\usepackage{comment}
\usepackage{csquotes}								% for nice looking quotation marks.
\usepackage{physics}
\usepackage{accents}

\def\bea {\begin{eqnarray}}
\def\eea {\end{eqnarray}}
\def\nn {\nonumber}
\def\p {\partial}

\begin{document}

\title{Anomaly in canonical semiclassical gravity}

\author{Viqar Husain} \email{vhusain@unb.ca}
\affiliation{Department of Mathematics and Statistics, University of New Brunswick, Fredericton, Canada}

\author{Irfan Javed} \email{i.javed@unb.ca}
\affiliation{Department of Mathematics and Statistics, University of New Brunswick, Fredericton, Canada}

\begin{abstract}

We show that the canonical formulation of the semiclassical Einstein equation, where the matter terms in the constraints are replaced by expectation values of the corresponding operators in quantum states, is inconsistent due to the non-closure of the resulting constraint algebra.   

\end{abstract}

\maketitle

\section{Introduction}
\label{intro}

Hybrid systems in which classical and quantum dynamics interact and influence one another are of interest from several perspectives, both in quantum mechanics and quantum field theory. One of the important questions in formulating hybrid models  \cite{Anderson:1995tn,Oppenheim:2018igd,Terno:2023hvq,Diosi:2024bsh} is whether it is possible to maintain the linear evolution of the wave function. 

The earliest approach is one where this does not hold, and the classical system couples with the expectation value of a quantum operator in an evolving state. This is illustrated by considering a state dependent hybrid Hamiltonian with classical variables $X,P$ and quantum ones $\hat{x},\hat{p}$: 
\bea 
H_\Psi = \frac{P^2}{2M} + V(X)  +  \langle\Psi| \frac{\hat{p}^2}{2m} + V(\hat{x})  + h_I (X,\hat{x}) |\Psi\rangle, 
\eea
with evolution given by Hamiltonian equations and the time dependent Schrodinger equation,
\bea 
\dot{X}&=& \{X, H_\Psi \},\ \ \dot{P}= \{P, H_\Psi \},\nn\\
i\frac{\p \Psi(x)}{\p t} &=& \left(\frac{\hat{p}^2}{2m} + V(\hat{x})  + h_I (X,\hat{x})\right)\Psi(x).
\label{hyb}
\eea
This type of hybrid model model includes full ``backreaction" of the quantum system on the classical one through the $\dot{P}$ equation, and of the classical system on the quantum system through the interaction term in (\ref{hyb}). The question of whether such an evolution can arise from a fully quantum dynamics of both systems, at least in some regimes, has been the subject of recent studies \cite{Husain:2022kaz, Husain:2023jaq}. The answer is that this can occur for a limited time only, as entanglement increases in a fully quantum bi-partite system, and initial semiclassical states do not remain so for long. 

Hybrid models in gauge theories require additional consideration. In non-Abelian gauge theory coupled to fermions, the interaction is of the form $g\bar{\Psi}\gamma^\mu A_\mu \Psi$. This leads to a source term in the Gauss law of the canonical theory
\bea 
D_a E^a = g\bar{\Psi}\Psi,
\eea
where $D_a = \partial_a + ig A_a$. In a hybrid theory where the gauge field is classical, the source term becomes an expectation value in a quantum state of the Fermi field, with no dependence on the gauge field. This is important because it avoids a gauge anomaly and the corresponding Hamiltonian remains gauge invariant, leading to a self-consistent hybrid theory. These considerations are important in applying  the hybrid scheme to gravity, where matter is quantum and gravity remains classical.   

The central feature of canonical formulations of general relativity is that the evolution of phase space variables is given by a linear combination of the spatial diffeomorphism and Hamiltonian constraints. These constraints satisfy a closed Poisson algebra, and so are themselves preserved under evolution. When matter is present, the phase space is extended to include the corresponding phase space variables, and the constraints are modified with additional terms corresponding to the energy density and flux of the matter field. These are the only changes, and the constraint algebra remains closed. 

In the Arnowitt-Deser-Misner  (ADM) canonical formulation \cite{Arnowitt:1962hi}, the phase space variables are the intrinsic 3-metric $q_{ab}$ of a spatial hypersurface embedded in the manifold, and its conjugate momentum $\pi^{ab}$ is a function of the extrinsic curvature of the hypersurface. The canonical action with a scalar field with phase space variables $(\phi,p_\phi)$
takes the form
\bea 
S &=& \int dt d^3x \left[\pi^{ab}\dot{q}_{ab} + p_\phi\dot{\phi}- N{\cal H} -N^a{\cal C}_a  \right], \\
{\cal H} &=& H^{\text G} + H^{\text M}, \\
{\cal C}_a &=& C_a^{\text G} + C_a^{\text M}, 
\eea
where $N$ and $N^a$ are the lapse and shift functions and the superscripts denote the gravity and matter components of the constraints. The corresponding spacetime metric is 
\bea 
ds^2 = -N^2dt^2 + (dx^a + N^a dt)(dx^b+ N^b dt)q_{ab}.
\eea
With the constraint functionals defined by ${\cal C}(N)=\int d^3x\ N^a {\cal C}_a$ and ${\cal H}(M) = \int d^3x\ M {\cal H}$, the Poisson algebra of the constraints is 
\bea 
\{ {\cal C}(N), {\cal C}(N') \} &=& {\cal C}([N,N']),\nn\\
\{ {\cal C}(N), {\cal H}(M) \} &=& {\cal H}({\cal L}_N M),\nn\\
\{{\cal H}(M), {\cal H}(M') \} &=& {\cal C}(W),
\label{c-alg}
\eea
where $[N,N']$ is the commutator of vector fields, ${\cal L}$ denotes the Lie derivative, and $W^a = q^{ab}(M\p_b M'- M'\p_b M ) $. 

The problem of canonical quantum gravity is that of representing this algebra as a commutator algebra on a suitably defined kinematical Hilbert space, solving the Dirac quantization conditions \cite{DeWitt:1967yk, carlip2015quantum}
\bea 
\hat{{\cal C}}(N) \Psi[q] =0, \quad \hat{{\cal H}}(M) \Psi[q] =0
\eea
to find the physical Hilbert space, and then studying the ``dynamics" of physical operators. This remains an unsolved problem both in the ADM variables, where it was originally posed, and in its later incarnation in the Ashtekar-Barbero connection-triad variables \cite{Thiemann:2001gmi,Ashtekar:2004eh,bodendorfer2016elementary}.

Given this impasse, one of the directions pursued is to seek suitably modified ``effective" constraints that are designed to capture desirable quantum gravity effects, such as fundamental discreteness and curvature singularity avoidance. There are many such proposals, and they accompany questions in addition to the required closure of the constraint algebra, such as the spacetime interpretation of the modified constraints --- see, e.g., \cite{Bojowald:2016hgh} and references therein. 

An apparently simpler problem is that of semiclassical gravity \cite{rosenfeld1963quantization,Ford:2005qz}, where matter is quantum and the gravitational field is classical: the postulated and much studied equation is 
\bea 
G_{ab} +\Lambda g_{ab} = 8\pi G\bra{\psi}T_{ab}(g, \hat{\phi}\ket{\psi}.
\label{sc}
\eea 
The quantum state $|\psi\rangle$ is given in the Heisenberg representation and the equation determines the ``semiclassical metric" $g$ given the stress-energy tensor operator. This leads to the immediate issue that $\hat{T}_{ab}(g)$ is also not known since it is a function of the unknown $g$. (In quantum field theory on curved spacetime, the metric $g$ is fixed and $\hat{T}_{ab}(g)$ can be well defined). There are also questions of interpretation,  such as the meaning of the metric for a state that is a linear combination of states that describe matter in disparate spacetime regions \cite{isham1995structural}. The equation may also be viewed as a modified theory of gravity with no reference to quantum states, where the right-hand side is a divergence-free rank two tensor made from the metric. 

\section{Canonical semiclassical gravity}

The question we address here is whether it is possible to define a self-consistent canonical version of (\ref{sc}).  The proposal we consider has the following  assumptions: (i) The quantum matter state is 
\bea 
\Psi[\phi,q] = \int d^3x\ \sqrt{q}\ g(x) f(\phi);
\label{mstate}
\eea
it is a functional of the spatial metric $q_{ab}$, the field $\phi$, and a smearing function $g(x)$ that may be used to localize points; it evolves via the functional Schrodinger equation (see e.g. \cite{Long:1996wf})
\bea 
\frac{\p}{\p t}\Psi[\phi] = \left(\hat{H}^{\text M}(M) + \hat{C}^{\text M}(N)\right) \Psi[\phi],
\label{funcSE}
\eea 
where the $t$ is coordinate time; 
(ii) The effective constraints (at each space point) are defined by 
\bea
{\cal H}^{\rm eff} &=& H^{\text G}(q,\pi) + \langle \hat{H}^{\text M} \rangle(q), \label{effH}\\
{\cal C}_a^{\rm eff} &=& C^{\text G}_a(q,\pi) + \langle \hat{C}^{\text M}_a \rangle(q),
\label{effC}
\eea 
where the expectation values are in the matter state (\ref{mstate}), and hence depend on the spatial metric;  (iii) The evolution of the gravitational variables is given by the constraint functionals $ {\cal H}^{\rm eff}(M) = \int d^3x \ M {\cal H}^{\rm eff}$ and   ${\cal C}^{\rm eff}(N) = \int d^3x \ N {\cal C}^{\rm eff}$:
\bea 
\dot{q}_{ab} &=& \{ q_{ab}, H^{\text G}(M) + C^{\text G}(N) \},\\
\dot{\pi}^{ab} &=& \{\pi^{ab}, {\cal H}^{\rm eff}(M) + {\cal C}^{\rm eff}(N) \}.
\label{geom_eqns}
\eea 
 The first of these equations is exactly the same as that for classical gravity since the expectation values of the matter terms in the effective constraints are functions of only the metric, and so do not modify the equation; for the same reason, the second equation is directly subject to backreaction due to the matter term expectation values.  

The question now is whether these equations are self-consistent; specifically, are the constraints preserved in time, or equivalently, does the algebra of effective constraints close? For homogeneous cosmologies, the spatial diffeomorphism constraint vanishes identically, and the evolution of the matter state and gravity variables is self-consistent \cite{husain2021quantum,husain2024semiclassical}. As we shall see, this is not the case for the inhomogeneous case.   

For illustration, consider the minimally coupled scalar field, for which the contributions to the Hamiltonian and diffeomorphism constraints are  
\bea
H^M &=&  \frac{p_\phi^2}{2\sqrt{q}} + \frac{\sqrt{q}}{2}q^{ab}\p_a\phi\p_b\phi  + \sqrt{q}V(\phi),  \nn\\
C^M_a &=& p_\phi\p_a\phi.
\eea
The expectation values in the effective constraints (\ref{effH}) and (\ref{effC}) are formally the functional integrals
\bea 
\langle \hat{H}^{\text M} \rangle &=& \int D\phi\ \bar{\Psi} [\phi] \left\{ -
\frac{1}{2\sqrt{q}}\frac{\delta^2 \Psi[\phi]}{\delta \phi(x)\delta \phi(x)} \right. \nn\\
&& \left.+ \sqrt{q}\left(\frac{q^{ab}}{2}\p_a\phi\p_b\phi  + V(\phi)\right) \Psi[\phi] \right\}  \nn\\
&=:& h(x,q(x)), \label{expecH}\\
\langle \hat{C}^{\text M}_a \rangle &=& -i\int D\phi\ \bar{\Psi}[\phi]  
 \frac{\delta \Psi[\phi]}{\delta \phi(x)} \p_a\phi(x)\nn\\
 &=:& c_a(x,q(x)).
 \label{expecC}
\eea
The space dependence in these expressions arises from the expectation value of the local operators at the point $x$, and through $q_{ab}(x)$; the latter dependence arises from both the operator and the state (\ref{mstate}) in the first case, and only through the state in (\ref{expecC}) in the second. The effective constraint functionals are then
\bea
{\cal H}^{\rm eff}(M) &=& \int d^3x\ M\left( H^{\rm G}(q,\pi) + h(x,q)  \right),\\
{\cal C}^{\rm eff}(N) &=& \int d^3x\ N^a\left(C^{\rm G}_a(q,\pi) +c_a(x,q) \right).
\eea 

\subsection{Constraint algebra anomaly}

For the diffeomorphism constraint brackets, let us make the assumption that the matter expectation values factor as 
\bea 
h(x,q(x)) &=& f(q(x))\alpha(x),\nn\\
c_a(x,q(x)) &=& g(q(x))\beta_a(x). 
\eea
This is reasonable because the explicit position dependence arises from their dependence on the matter operators, and it also serves to highlight the effect on the algebra. It follows that   
\bea
  \{c(N), C^G(N') \} &=& c({\cal L}_{N'}N) \nn\\
  && + \int d^3x\ g(q)N^a{\cal L}_{N'} \beta_a, \\
 \{h(M), C^G(N)\} &=& h({\cal L}_{N}M) \nn\\
 && + \int d^3x\ f(q) M {\cal L}_{N} \alpha.
\eea
The Poisson brackets involving ${\cal C}^{\rm eff}$ are 
\bea 
 \{{\cal C}^{\rm eff}(N),{\cal C}^{\rm eff}(N') \} 
 &=& C^{\rm G}([N,N']) 
  +\ \{C^{\rm G}(N),c(N')\} \nn\\
  && + \{ c(N),C^{\rm G}(N')\}\nn \\\nn\\
 &=& {\cal C}^{\rm eff}([N,N']) \nn\\
 &&+ \int d^3x\ g(q) [N,N']^a\beta_a,
 \label{DiffAnom}
 \eea
 \bea
 \{ {\cal H}^{\rm eff}(M), {\cal C}^{\rm eff}(N) \} &=&  
 \{ H^{\text G}(M) + h(M) ,C^{\text G}(N)  \}\nn\\
 && +\ \{ H^{\text G}(M) + h(M),c(N) \}\nn\\
 &=& {\cal H}^{\rm eff}({\cal L}_N M) + \{ H^{\text G}(M),c(N) \}\nn\\
&& + \int d^3x\ f(q) M {\cal L}_N \alpha.
\label{HeffAnom}
\eea
The remaining bracket is  
\bea 
\{ {\cal H}^{\rm eff}(M), {\cal H}^{\rm eff}(M')\}&=& \{ H^{\rm G}(M), H^{\rm G}(M')\} \nn\\
&=& C^{\rm G}(W)\nn\\
&=& {\cal C}^{\rm eff}(W) -c(W),
\eea
where $W^a$  is the same vector field as in the classical algebra (\ref{c-alg}). Thus, every bracket has an anomaly that depends in general on the spatial metric and the matter term expectation values---this is the main result of this note. 

\section{Discussion}

We defined a canonical semiclassical approximation for general relativity coupled to matter fields, using the same general idea as that for the covariant semiclassical Einstein equation (\ref{sc}), namely that matter terms are replaced by their expectation values in quantum states. This leads to defining effective semiclassical constraints (\ref{effH}) and (\ref{effC}), and the  generation of evolution equations from them. As we have shown, this leads to the inconsistency that the algebra of effective constraints does not close. This means that the spacetime reparametrization invariance does not follow from the canonical theory as it does in classical general relativity \cite{Hojman:1976vp}. Our result may be viewed as further ``indirect evidence" for the quantization of gravity \cite{page1981indirect}.  

A similar anomaly follows if the starting point for postulating the semiclassical equations is modified by imposing a time gauge to determine a physical Hamiltonian. One such case is the dust time gauge \cite{Brown:1994py,Husain:2011tk}, where the physical Hamiltonian takes the same expression as the Hamiltonian constraint (\ref{effC}) except that it is no longer constrained to be zero. Only the effective diffeomorphism constraint remains; it is then immediate that the constraint algebra is still (\ref{DiffAnom}), and the physical Hamiltonian is not diffeomorphism invariant, a fact that follows from (\ref{HeffAnom}). Thus, no time gauge fixing will resolve the anomaly problem, since the anomaly in the spatial diffeomorphism algebra persists.

It is possible to consider a different semiclassical approximation where the equations of motion are not derived from effective constraints. Instead, the canonical equations of motion are first derived from the fully classical constraints and then modified by hand to replace all matter terms by their expectation values, supplemented with the functional Schrodinger equation (\ref{funcSE}). This approach is closer to the original covariant proposal (\ref{sc}). Static black hole and boson star solutions in spherical symmetry have been derived using this alternative approach by considering stationary matter states \cite{Javed:2025ogr}.  

In summary, we have shown that the canonical version of the semiclassical approximation for gravity is not self-consistent, unlike the situation in gauge theory.    

\medskip
\noindent {\bf Acknowledgements} This work was supported by the Natural Science and Engineering Research Council of Canada.

\bibliography{ref}
\end{document}